\documentstyle[aps]{revtex}

\begin{document}
\draft
\title{Distribution of the wavefunction inside chaotic partially open systems}
\author{P. {\v S}eba}
\address{Nuclear Physics Institute, Czech Academy of Sciences, 250 68
\v Re\v z, Czech Republic\\
Pedagogical University, Hradec Kralove, Czech Republic}
\author{F. Haake}
\address{Fachbereich Physik, Universit\"at-GH Essen, Essen, Germany}
\author{M. Ku\'s}
\address{Center for Theoretical Physics, Polish Academy of Sciences, Warsaw, Poland}
\author{M. Barth \and U. Kuhl \and H.-J. St\"ockmann}
\address{Fachbereich Physik, Universit\"at Marburg, D-35032 Marburg, Germany}

\date{\today}

\maketitle

\begin{abstract}
We demonstrate both theoretically and experimentally that the 
distribution of the wavefunction inside a partially open chaotic 
timereversal symmetric system displays significant deviations from the Porter
Thomas distribution. We give arguments which show that this distribution
resembles the distribution which is expected to be found in closed chaotic
systems with broken time reversal symmetry.
\end{abstract}


It is now generally accepted that in the ballistic regime the scattering on
boundaries of a chaotic mesoscopic quantum dot leads to irregular scattering
phenomena which are measurable during the transport of electrons through the
dot. Usually it is assumed that ideal leads are attached to the quantum dot
and the conductance $G$ is evaluated using Landauer formula relating the
conductance with the corresponding S matrix
\begin{equation}
S=\left(
\begin{array}{cc}
r&t\\
t'&r'\\
\end{array}
\right)
\end{equation}
where $r,t$ are the reflection and transmission matrices. In terms of S
matrix the conductance reads
\begin{equation}
G=\frac{e^2}{h} \textrm{Tr} \left(tt^+\right).
\end{equation}

Essentially the same mechanism can be applied also to the transmission of
microwaves through irregularly shaped cavities. In this case $G$ denotes the
total transmission probability (of course in this case the factor $e^2/h$
has to be omitted!). The microwave experiments have the advantage that all
components of the scattering matrix are obtained directly from the
measurement, and that scattering geometries can be easily varied in a
controlled manner. Therefore the results to be derived below will be tested
using results from microwave cavities containing randomly distributed
scatterers.

Investigating conductance fluctuations the standard argument assumes that
the S matrix belongs to some random matrix ensemble (usually the circular
orthogonal (COE) or unitary (CUE) ensemble). There is, however, also another
aspect of the statistical properties of the S matrix elements which relates
the S matrix elements with the internal wavefunction.

The S matrix maps the incoming waves into the outgoing ones. Therefore
knowing S and the structure of the incoming part of the wave inside the
leads one can easily evaluate the value of the wavefunction at the points
where the leads couple to the resonator. Since we assume an ideal coupling
of the leads on the quantum dot the wavefunction is smooth at the coupling
points. This means that the internal wavefunction at {\it the coupling
points} equals the wavefunction inside the leads. Therefore from a knowledge
of the statistical properties of the S matrix we can obtain information
about the properties of the internal wavefunction.

It is the aim of this letter to develop the above heuristic arguments and to
show to what extent the transport through the dot changes the structure of
the wavefunction. Let us start with some remarks: the important role of the
internal wavefunction during the transport through weakly open systems
(resonance transport) has been used in \cite{pri93,alh97a} to investigate
the statistical properties of the conductance. In these papers the applied
strategy was however just opposite to what we try to do here: the
conductance distribution was obtained {\it assuming} that the internal
wavefunction is chaotic and has a Porter-Thomas distribution. On the other
hand there are also works which followed the same strategy and used the
measured S matrix to determine the internal wavefunction, see for instance 
\cite{alt95,ste95}. It has to be stressed, however, that in all 
these cases theinvestigations are restricted to the resonance scattering 
regime and theinfluence of the transport on the structure of the internal 
wavefunction hasbeen neglected.

To begin the investigation let us first construct a simple Hamiltonian
which leads to the S matrix being usually used as a starting point for the
description of transmission fluctuations. Starting with the description of
the leads we assume that they support $M$ open channels and are described by
one-dimensional Hamiltonians
\begin{equation}
H_l = - \frac{d^2}{dx^2} + \lambda_l \ , \ l=1,..,M .
\end{equation}
Here $\lambda_l$ is the threshold energy of the $l$-th channel and $x$
denotes the coordinate along it. Combining these operators into a
Hamiltonian $H_{ex}$ for the ``external'' part of the system we get
\begin{equation}
\label{Schroedinger} 
H_{ex} = -{\bf 1} \frac{d^2}{dx^2} + \Lambda,
\end{equation}
where $\Lambda$ is a diagonal matrix describing the threshold energies of
the channels,
\begin{equation}
\Lambda = diag\left( \lambda_1, \lambda_2,..,\lambda_M \right),
\end{equation}
and $1$ is the the $M \times M$ identity matrix. The resonator is described
by a Hermitian matrix $H_{in}$ of size $N \times N$, where $N$ corresponds
to the number of eigenenergies taken into account, with $N$ much larger than
the number of open channels, $N \gg M$. The matrix $H_{in}$ is assumed to
belong to the Gaussian orthogonal / unitary Ensemble (GOE/GUE) for chaotic
cavities (see e.g. \cite{haa91a} for these concepts). To describe the
scattering we couple the resonator and leads by defining the Hamiltonian 
$H$of the whole system as
\begin{equation}
\label{hamiltonian}
H\left(
\begin{array}{ll}
u\\
u_{in}
\end{array}
\right)
= \left(
\begin{array}{cc}
H_{ex} u   \\
H_{in} u_{in}+ Au'(0)
\end{array}
\right),
\end{equation}
where $u = (u_1,\ldots,u_M)$ stands for the wavefunction inside the leads
and $u_{in}$ describes the wavefunction within the resonator. $u'(0)$
denotes the vector of derivatives of the wave functions inside the leads
taken at the  points of contacts with the resonator (i.e.\ at zero of each
lead coordinate). $A$ is the $N\times M$ coupling matrix.
Later on we will specify the coupling matrix $A$ by assuming that the
coupling has a local character (point contacts). This assumption is
justified whenever the diameter of the junction is smaller than a typical
wavelength inside the resonator.

Let us now return to the Hamiltonian $H$. In the form it is given by
(\ref{hamiltonian}) the operator is not symmetric. To make it symmetric an
additional boundary condition is needed:
\begin{equation}
\label{boundary}
A^\dagger u_{in} = -u (0).
\label{bound}
\end{equation}
It is not difficult to show, that under these condition the Hamiltonian $H$
is a self-adjoint operator -- for the proof see \cite{alb96}.

Before proceeding further we have to specify the structure of the internal
Hamiltonian matrix $H_{in}$. On the most general level we assume only that
this matrix belongs to the Gaussian orthogonal (unitary) ensemble. Less
abstractly it will be however helpful to construct this matrix using the
knowledge of the specific properties of the resonator in consideration. To
specify $H_{in}$ we use the Hamiltonian $H_{res}$ of the 
resonator (usually a two dimensional Laplace operator with Dirichlet 
boundary conditions). Let $E_n$ and $f_n(\vec r)$ be the 
eigenvalues and eigenfunctions of $H_{res}$, 
\begin{equation} 
H_{res} f_n(\vec r) = E_n f_n(\vec r).
\end{equation}

Using these solutions we define a finite dimensional internal Hamiltonian
acting on the space spanned by the first $N$ eigenstates of $H_{res}$,
\begin{equation}
H_{in} = \sum_{n=1}^N E_n f_n f_n^{\dagger}.
\end{equation}

The coupling operator $A$ maps the vector $u'(0)$ into a certain function
belonging to the $N$ dimensional internal space. Let $ u(x) = \left(u_1(x),
u_2(x),\ldots,u_M(x)\right) $ denote the components of the wave function in
the attached open channels. Applying the matrix $A$ to the incoming vector
we get
\begin{eqnarray}
\label{incoming}
A u'(0) &=&\sum_{m=1}^M\sum_{n=1}^N u_m'(0)A_{nm}f_n \nonumber \\ &=&
\alpha_1 d^N_1 (\vec r) u_1'(0) +\alpha_2 d^N_2 (\vec r)
u_2'(0)+\ldots+\alpha_M d^N_M (\vec r) u_M'(0),
\end{eqnarray}
where $\alpha_ld_l^N=\sum_{n=1}^NA_{nl}f_l$;  $\alpha_1$, \ldots, $\alpha_M$
are the coupling constants of the individual channels and $ d_l^N, \;
l=1,\ldots,M $ are functions spanned by the vectors $f_n; \; n=1,\ldots,N$.
In the experiment the channels are locally coupled to the resonator at
points $\vec r_1$, \ldots, $\vec r_M$. Here $\vec r_1$, \ldots, $\vec r_M$
refer to the coordinate system chosen inside the resonator and correspond to
the zero points of the coordinates inside the leads. In order to mimic this
local coupling we choose the functions $ d_l^N(\vec r),\; l=1,\ldots,M $ in
a special way which ensures their convergence to $\delta (\vec r -
\vec r_l)$ for $N\to\infty$, namely
\begin{equation}
\label{haake}
d_l^N(\vec r) = \sum_{i=1}^N f_i(\vec r_l) f_i(\vec r).
\end{equation}
In the sense of generalized functions one has $\lim_{N\to\infty} d_l^N(\vec
r) = \delta (\vec r-\vec r_l)$.

At this point few remarks are necessary: First of all it has to be stressed
that now the coupling matrix $A$ is completely determined by $M$ constants
$\alpha_1$, \ldots, $\alpha_M$. The boundary condition (\ref{bound}) takes
now the form $ <\alpha_l d_l^N, u_{in}> = -u_l (0) $, where $<,>$ denotes
the scalar product in the $N$-dimensional Hilbert space of functions inside
the resonator. In the limit $N\to\infty$ this boundary conditions are thus
given by the formula
\begin{equation}
\label{relation}
\alpha_l u_{in} (\vec r_l) = - u_l(0),
\end{equation}
which relates the internal wavefunction at the coupling point to the
wavefunction inside the corresponding lead. If we assume that the contact is
ideal, i.e.\ that the wavefunctions match smoothly at the contact, $u_{in}
(\vec r_l) =  u_l (0)$,  we obtain $\alpha_l=-1$ for all $l$. The components
of the coupling matrix $A$ are then given by
\begin{equation}
\label{coupl}
A_{lm}=-f_l(\vec r_m).
\end{equation}

Equation (\ref{relation}) becomes a starting point for a further
consideration of the statistical properties of the internal wavefunction.

Since the information about the structure of the internal wavefunction is
hidden in the scattering data we have first to evaluate the corresponding S
matrix. Following \cite{alb96} we solve the equation $HU=EU$ for a
scattering energy $E$. The eigenfunction
\begin{equation}
U(E) = \left(\begin{array}{c}u (E,x) \\ u_{in} \end{array} \right)
\end{equation}
solves the equations
\begin{equation}
\label{second}
\left(
\begin{array}{c}
- \frac{d^2}{dx^2} u (E,x) + \Lambda u (E,x) \\
H_{in} u_{in} (E) + A u' (E,0) \end{array} \right)
= E \left(
\begin{array}{c}
u (E,x) \\
u_{in} \end{array} \right).
\end{equation}

For energies $E > \lambda_j,\ j=1,\ldots,M$, the scattering solution inside
the leads can be presented in the form
\begin{equation}
\label{norma}
u (E,x) =
\frac{ e^{-i \sqrt{ E- \Lambda} x}}{\root 4 \of {E-\Lambda}}{\cal A}_{inc}
-
\frac{ e^{i \sqrt{ E- \Lambda} x}}{\root 4 \of {E-\Lambda}}{\cal A}_{out},
\end{equation}
where ${\cal A}_{inc}, {\cal A}_{out}$ are the amplitudes of the incoming
and outgoing wave, respectively. For all $E > max \{ \lambda_j\}$ every
solution is bounded and the scattering matrix $S(E)$ can easily be defined.
The normalization used in (\ref{norma}) ensures that the S matrix relates
the amplitudes of the incoming and outgoing waves as
\begin{equation}
{\cal A}_{out} = S(E) {\cal A}_{inc}.
\end{equation}

The scattering matrix can be calculated substituting (\ref{norma})
into (\ref{second}) and applying the boundary condition (\ref{boundary})
which leads to
\begin{equation}
\label{smatrix}
S(E) = \frac{ i + W^\dagger (E-H_{in})^{-1} W }
{ i - W^\dagger (E-H_{in})^{-1}W },
\end{equation}
where $W=AQ^{-1}$ and $Q \equiv Q(E)$ denotes the $M\times M$ matrix
$ Q(E) = (E-\Lambda)^{-\frac{1}{4}}$. This S matrix can be rewritten as
\cite{alb96}
\begin{equation}
S(E)=1-2i  W^\dagger \frac{1}{E-H_{eff}}W,
\end{equation}
where $H_{eff}=H_{in}+i WW^\dagger$ stands for the effective, nonhermitian
Hamiltonian. The assumption on the point character of the contact leads in
the limit $N\to\infty$ to $H_{eff} = H_{res} + i \sum_{n=1}^M
\alpha_n^2\sqrt{E-\lambda_n} \ \delta(\vec r-\vec r_n)$. Effective
Hamiltonians of this type has been used previously on a heuristic level in
\cite{pri93} for the description of conductance fluctuations in quantum
dots. In the present case, however, the coupling of the imaginary effective
potential depends on energy (frequency) of the incident wave. This fact has
been observed experimentally in \cite{haa91b}.

For the value of the external component
$u(E,x)$ at the coupling points we get from (\ref{incoming})
\begin{equation}
u(E,0) = Q (1-S) {\cal A}_{inc};
\end{equation}
and finally from (\ref{relation}) a relation between the S matrix and
the internal function:
\begin{equation}
\label{basic}
\left(
\begin{array}{c}
u_{in}(\vec r_1) \\
u_{in}(\vec r_2) \\
\vdots \\
u_{in}(\vec r_M)
\end{array}
\right)
=  Q(S-1){\cal A}_{inc}.
\end{equation}

The relation (\ref{basic}) is valid not only for the case of a resonant
scattering but also in situations with many overlapping resonances. This can
be understood when the formula is compared with previous results, where a
relation between the resonance wavefunction and the S matrix was obtained
for the case of one {\it isolated} resonance \cite{pri93,alh97a,ste95}. The
relation (\ref{basic}) gives us the possibility to look for the structure of
the internal wavefunction in situations where the S matrix belongs to
certain ensembles of random unitary matrices.

In a typical experiment the wave is fed into the resonator through one of
the attached channels - say through the first one - and the wavefunction is
measured either by measuring the reflection in the entrance lead or the
transmission to an exit lead. This means that ${\cal A}_{inc}=(1,0,..,0)$
and
\begin{equation}
\label{function}
u_{in}(\vec r_k) =
\frac{1}{(E-\lambda_k)^{1/4}}\left(S_{1k} - \delta_{1k}\right)
\end{equation}
with $k=1$ in the reflection case and $k>1$ in the transmission case.
Assuming $\lambda_k$ to be constant (i.e. not varying from sample to sample)
we obtain finally that the statistics of $u_{in}(\vec r_k)$ coincides with
the statistics of the S matrix elements $S_{1k}$.

It is widely accepted that the structure of the S matrix of a chaotic system
with time reversal symmetry is described by a generalized circular ensemble
(so called Poisson kernel)
introduced by Gaudin and Mello \cite{gau81}, see also \cite{mel85}. The
structure of this ensemble is fully determined by the mean value $<S>$ of
the S matrix. The mean value describes the "quality" of the contact.  It is
related to the coupling matrix $W$ by \cite{ver85a}
\begin{equation}
<S_{ll}>=\frac{1-\gamma_l}{1+\gamma_l}
\end{equation}
with $\gamma_l$ given by
\begin{equation}
\label{orthogonal}
\sum_{n=1}^N W_{nl}W_{mn}=\delta_{lm}\gamma_m.
\end{equation}

Note that the orthogonality relation (\ref{orthogonal}) follows from
(\ref{coupl}) and is satisfied whenever the distance between the coupling
points exceeds the typical wavelength in the scattering problem. In the case
of ideal coupling we have $<S>=0$ and the ensemble coincides with the
Circular Orthogonal Ensemble of Dyson. The relation between the Poisson
kernel and S matrices of the form (\ref{smatrix}) is described in
\cite{bro95b}.

The statistics of the corresponding S matrix elements $S_{kl}$ was evaluated
in \cite{per83}. It was shown that for sufficiently large $M$ the S-matrix
elements can be regarded as statistically independent. Moreover the real and
imaginary part of $S_{kl}$ are independent Gaussian distributed variables.

Let us first consider a measurement of the transmission between leads 1 and
2. Denoting $S_{12}=X+iY$ and following \cite{per83} we obtain
\begin{equation}
\label{ps12}
P(S_{12})=\frac{\sqrt{a b}}{\pi}e^{-aX^2} e^{-bY^2}
\end{equation}
with $a,b$ given by
\begin{equation}
\label{ab12}
\begin{array}{c}
a = \frac{1+<S{11}><S{22}>}{(1-<S_{11}>^2)(1-<S_{22}>^2)} \\ b =
\frac{1-<S{11}><S{22}>}{(1-<S_{11}>^2)(1-<S_{22}>^2)}
\end{array}
\end{equation}
and $<S_{11}>,<S_{22}>$ being the mean values of the corresponding matrix
elements (they are assumed to be real for simplicity). Finally using
(\ref{function}) we obtain for the distribution of the modulus of the
normalized internal wavefunction at point $\vec r_2$ \cite{kan96}
\begin{equation}
\label{distribution}
P\left(|u_{in}|^2\right) =
Z \exp\left(-Z^2|u_{in}|^2\right) I_0\left(Z\sqrt{Z^2-1}|u_{in}|^2 \right)
\end{equation}
with
\begin{equation}
\label{z12}
Z=\frac{1}{\sqrt{1-<S_{11}>^2<S_{22}>^2}}.
\end{equation}

A similar relation holds also for the reflection measurement.  The internal
wavefunction is studied using the matrix element $S_{11}$ which measures the
direct reflection of the wave into the incoming channel. For the
distribution of $S_{11}=X+iY$ one has \cite{per83}:
\begin{equation}
\label{ps11}
P(S_{11}) = \frac{\sqrt{a b}}{\pi}e^{-a(X-<X>)^2} e^{-bY^2}
\end{equation}
with the coefficients $a,b$ given by
\begin{equation}
\label{ab11}
\begin{array}{c}
a = \frac{1+<S{11}>^2}{(1-<S_{11}>^2)^2} \\ b = \frac{1}{1-<S_{11}>^2}
\end{array}
\end{equation}
and with $<S_{11}>=<X>$ denoting the mean of the S matrix element. Using
that we get for the distribution of the normalized internal wavefunction
\begin{equation}
<u_{in}> = 1 - <S_{11}>
\end{equation}
and for the distribution of $\tilde u_{in} = u_{in} - <u_{in}>$
\begin{equation}
\label{distribution2}
P\left(|\tilde u_{in}|^2\right) = Z \exp\left(-Z^2|\tilde u_{in}|^2\right)
I_0\left(Z\sqrt{Z^2-1}|\tilde u_{in}|^2 \right)
\end{equation}
with $Z$ equal to
\begin{equation}
\label{z11}
Z=\frac{1}{\sqrt{1-<S_{11}>^4}}.
\end{equation}

In the COE case with $<S>=0$ (ideal coupling of channels) both distributions
(\ref{distribution}) and (\ref{distribution2}) lead to Poissonian
distribution:
\begin{equation}
P(\left(|u_{in}|^2\right) = \exp\left(-|u_{in}|^2\right).
\end{equation}

In the other extreme case of very weakly coupled channels for which we have
$<S_{11}> \approx <S_{22}>\approx \pm 1$ and hence $Z\to\infty$, which
reduces (\ref{distribution}) to the Porter-Thomas-distribution
\begin{equation}
P\left(|u_{in}|^2\right) =
\frac{1}{\sqrt{2\pi |u_{in}|^2}}\exp\left(-\frac{|u_{in}|^2}{2}\right).
\end{equation}

The same holds also for the distribution (\ref{distribution2}) with
$<S_{11}>\approx \pm 1$.

It is worth stressing that in the COE case the distribution of the internal
wavefunction coincides with the distribution of eigenvectors of an GUE
ensemble, i.e. with a case of fully broken time reversal symmetry.

For smaller number of channels the distribution of the $S$-matrix elements
does not factorize and the distribution  of the internal wavefunction cannot
be  found explicitly. Nevertheless it becomes clear that it depends strongly
on the number of open channels. For instance in the case of two ideally
coupled channels (COE case) we obtain from \cite{bar94}
\begin{equation}
P\left(|u_{in}|^2\right) = \frac{1}{2\sqrt{2\pi |u_{in}|^2}}.
\end{equation}

Microwave billiards are especially well suited to test the predictions on 
distributions of the $S$-matrix elements (seeEqs.~(\ref{ps12}) and 
(\ref{ps11})) and of transmission and reflectionprobabilities (see 
Eqs.~(\ref{distribution}) and (\ref{distribution2})). Allquantities entering 
into these expressions are directly available from theexperiment. It is of 
special importance that there are only two parameters,namely the averages 
$<S_{11}>$ and $<S_{22}>$. As these quantities, too, canbe measured, there 
is no free adjustable parameter in the theory. 

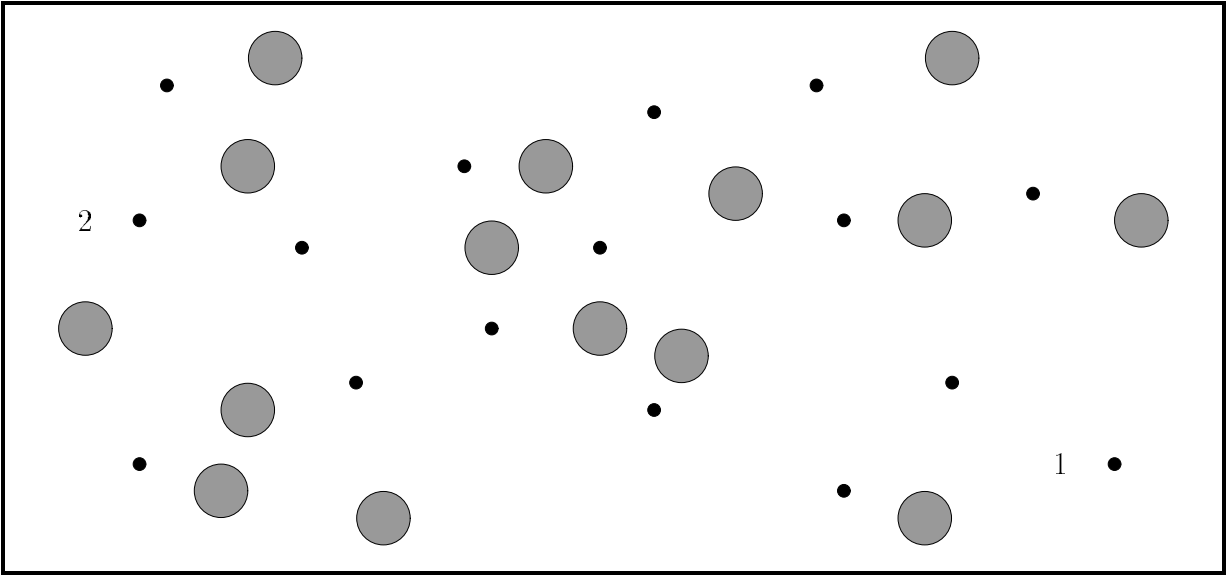
\begin{figure}
\caption{Rectangular billiard (45 cm $*$ 20 cm, height 0.8 cm) with 13 to 17
movable scatterers of diameter 2 cm positioned at random on a 18$*$8-grid.
There are 16 fixed antennas (copper wires of diameter 0.2 mm) connected to
the billiard at randomly chosen points with a minimum distance of 2.5 cm.}
\label{billiard}
\end{figure}

The measurements were performed in a rectangular microwave resonator with 10
to 17 randomly distributed scatterers; fig.~\ref{billiard} shows a typical
arrangement. The number of scatterers should be sufficient to block most of
the bouncing ball modes in order to make the system chaotic; on the other
hand the mean distance should be at least of the order of the typical
wavelength. The measuring technique was described earlier
\cite{stoe90,ste95}. A vector network analyzer, model 360B, Wiltron company,
was used supplying real and imaginary part of all components of the S
matrix.

To study the transport through the resonator 16 antennae (thin copper wires
of diameter 0.2 mm) were put into the resonator. The antennae act as single
scattering channels as their diameter is small compared to the wavelength in
the total frequency range. Only two antennae were really used for reflection
and transmission measurements. Through the antenna 1 microwaves are coupled
to the resonator and reflection ($S_{11}$) is measured. The transmission
($S_{12}$) is measured with the help of the antenna 2 . All other antennae
not used are closed by 50 $\Omega$ loads and act as drains for the
microwaves.

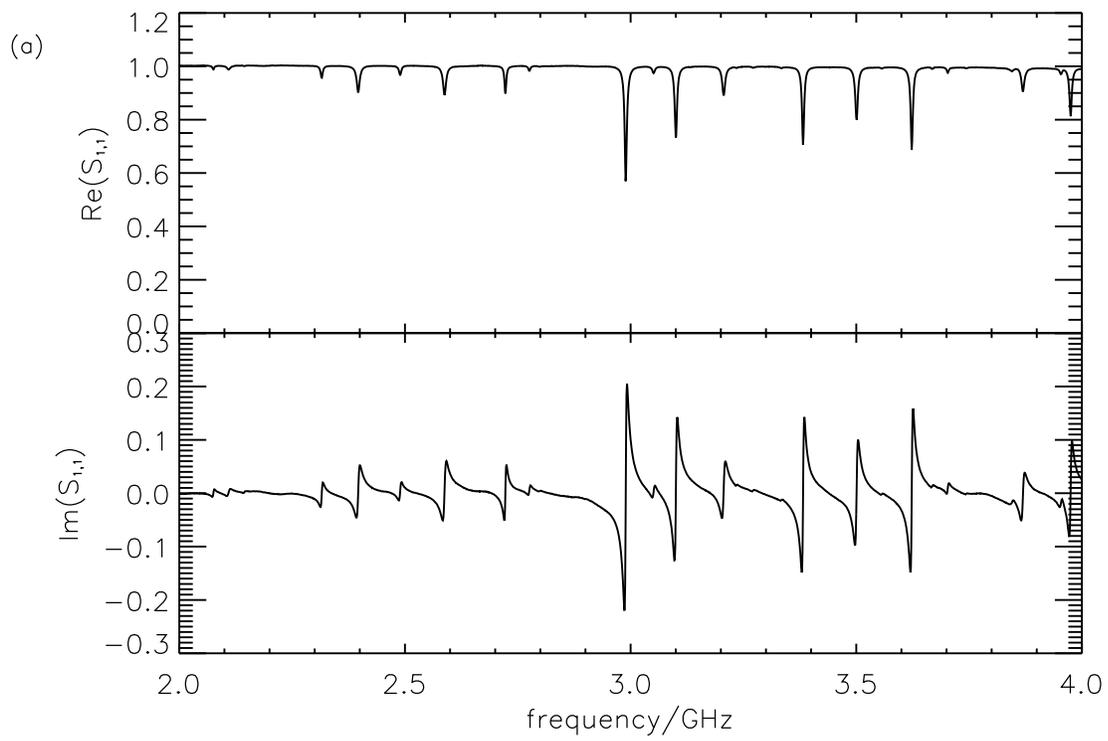
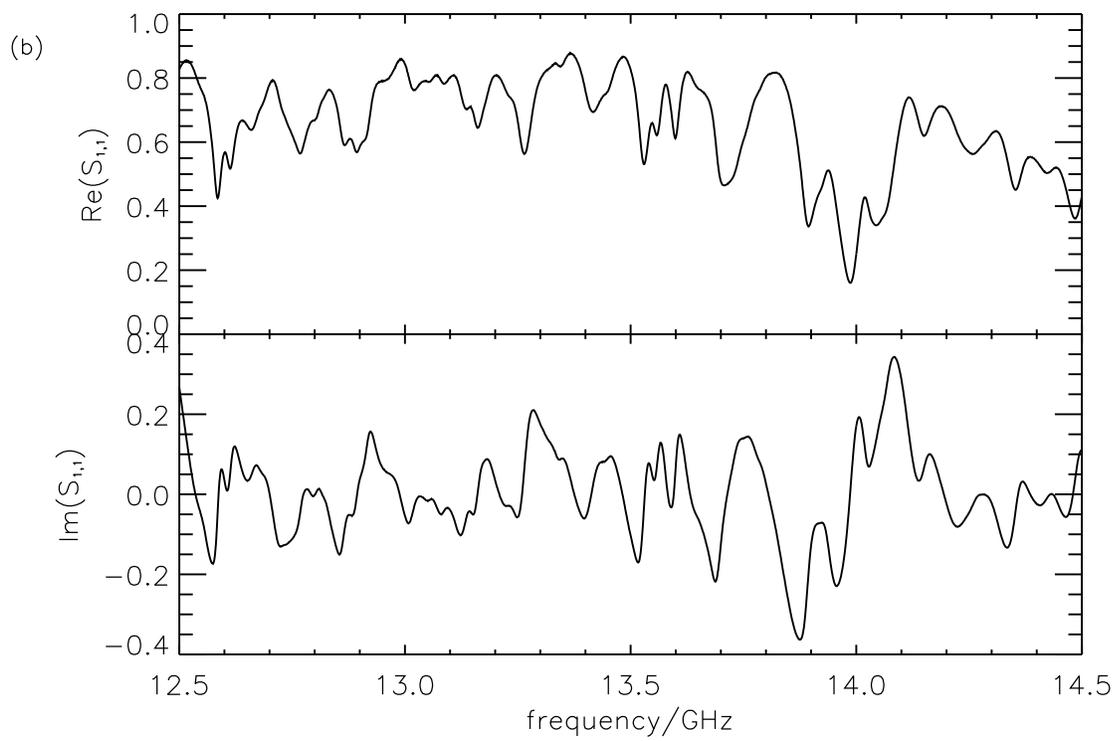
\begin{figure}
\caption{Parts of a typical $S_{11}$ spectrum after calibrating away the
effects of cables and connectors. At lower frequencies the resonances are
sharply separated whereas at higher frequencies they overlap.}
\label{spectra}
\end{figure}

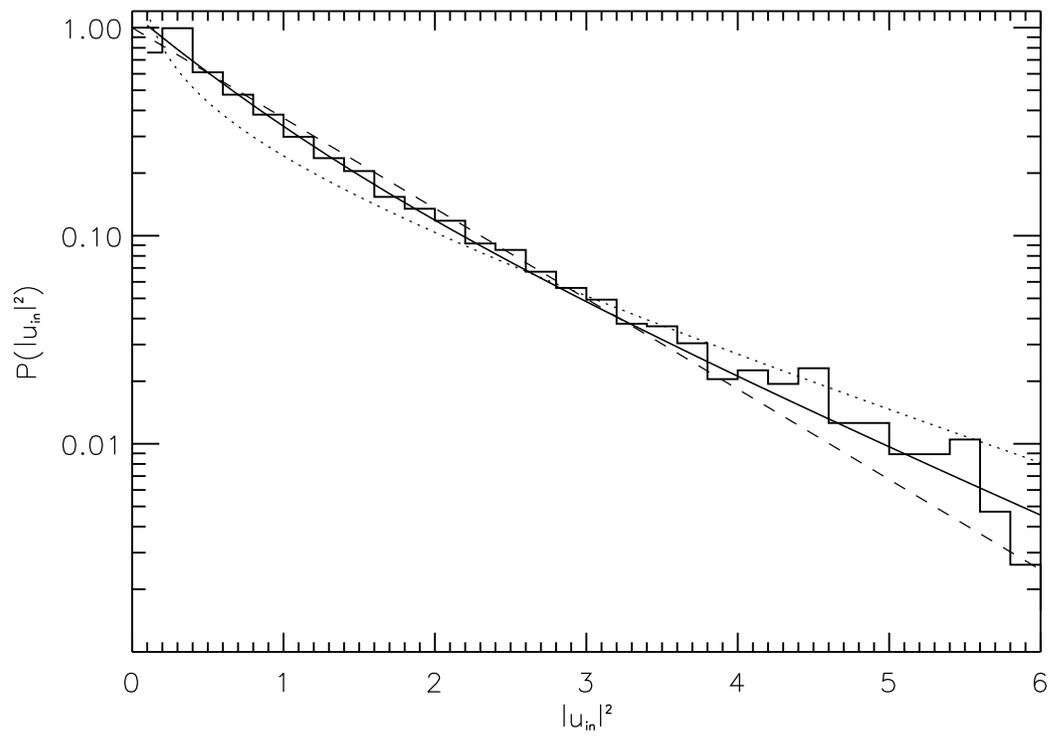
\begin{figure}
\caption{Histogram of the $|u_{in}|^2$-values at low frequencies (non
overlapping regime) in a semilogarithmic plot. The dashed and dotted lines
correspond to Poisson distribution ($Z=1$, GUE) and
Porter-Thomas distribution ($Z\to\infty$, GOE), respectively. The solid line
depicts the distribution (\ref{distribution2}) with the parameter $Z\approx
1.17$ calculated from $<|u_{in}|^2>$.}
\label{low}
\end{figure}

In Fig.~\ref{spectra} parts of a typical measured $S_{11}$-spectrum are
shown. The qualitative difference between the non-overlapping regime at
lower frequencies (a) and the overlapping regime at higher frequencies (b)
is immediately evident. For the case of non-overlapping resonances we
performed about 600 measurements with different positions of ten 
scatterers from 2 to 4 GHz with a resolution of 1 MHz. In the region of well
separated resonances one can derive from the billiard equivalent of the
Breit-Wigner formula a relation between $|u_{in}|^2$ and the measured
$S_{11}$ at the maximum of the resonances,
\begin{equation}
|u_{in}|^2 \propto \Gamma (1 - {\textrm Re}(S_{11})),
\end{equation}
where $\Gamma$ is the width of the resonance \cite{ste95}. We fitted the
single resonances in the real part of S$_{11}$ with a Lorentzian and
extracted height and width of them to obtain $|u_{in}|^2$. In total we
obtained a sample of more than 12000 values for $|u_{in}|^2$. The histogram
of these values is shown in Fig.~\ref{low} and compared to the theoretical
curves for GOE and GUE and the distribution (\ref{distribution2}) using $Z
\approx 1.17$, calculated from the experimentally obtained $<|u_{in}|^2>
\approx 0.72$. The experimental distribution fits very well to the
theoretical one. It should be noted that no parameter had to be fitted to
obtain this accordance.

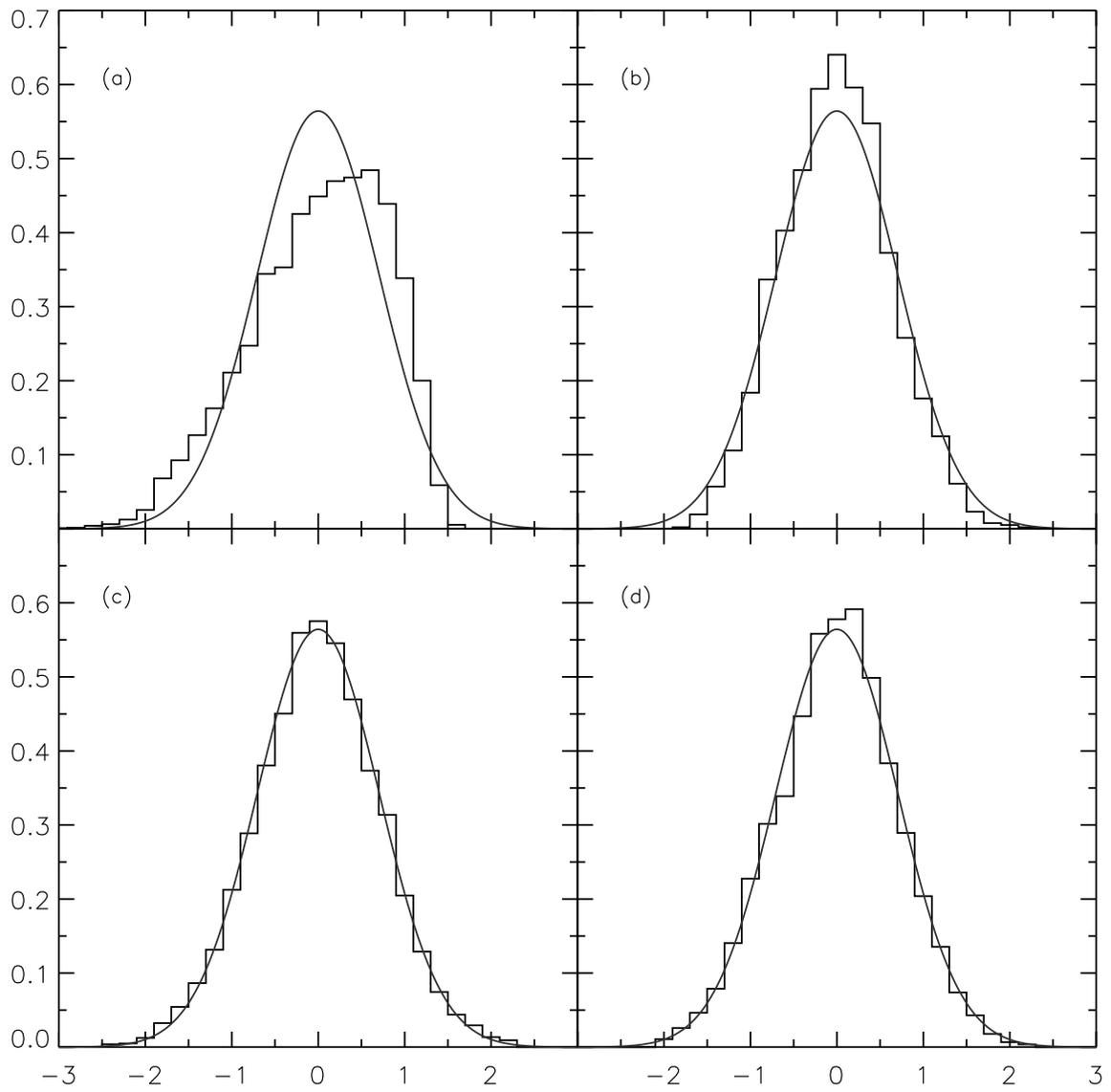
\begin{figure}
\caption{The distributions of real and imaginary parts of $S_{11} - <S_{11}>$
(a,b) and $S_{12}$ (c,d). All distributions are normalized to have the
variance equal to one. The solid lines correspond to the expected Gaussian
behaviour.}
\label{gauss}
\end{figure}

At higher frequencies where single resonances are no longer well resolved we
can take the total spectrum for the determination of the S-parameters. The
calibration poses a problem. By application of standard procedures the
influence of cables, connectors etc.\ is efficiently calibrated away. The
influence of the antenna wire itself, however, cannot be removed by the
calibration and results in a long range variation of baseline and phase over
several GHz. Apart from some regimes, which were excluded from the further
analysis, the drift of the phase could be corrected away by a polynomial
background subtraction. It was not possible, however, to discriminate
between phase drifts from the antenna and real phase shifts by the billiard.
Therefore the determination of the phase of $S_{11}$ is necessarily
erroneous. This can be seen in Fig.~\ref{gauss}a,b where the distributions
of ${\rm Re}(S_{11}-<S_{11}>)$ (a) and of ${\rm Im}(S_{11})$ (b) are
plotted. The error in the phase determination is responsible for 
the clear deviation of the distribution of ${\rm Re}(S_{11}-<S_{11}>)$ from 
a Gaussian. For ${\rm Im}(S_{11})$, on the other hand, exactly the expected 
Gaussian behavior is found though here of course, too, the phase 
determination is incorrect. But as the average of ${\rm Im}(S_{11})$ 
vanishes in contrast to that of ${\rm Re}(S_{11})$, an error in the phase 
determination is not able to disturb the Gaussian behaviour. It should be 
noted that errors in the phase do influence only the distribution of ${\rm 
Re}(S_{11} - <S_{11}>)$, for all other distributions discussed here this 
phase is not of relevance.

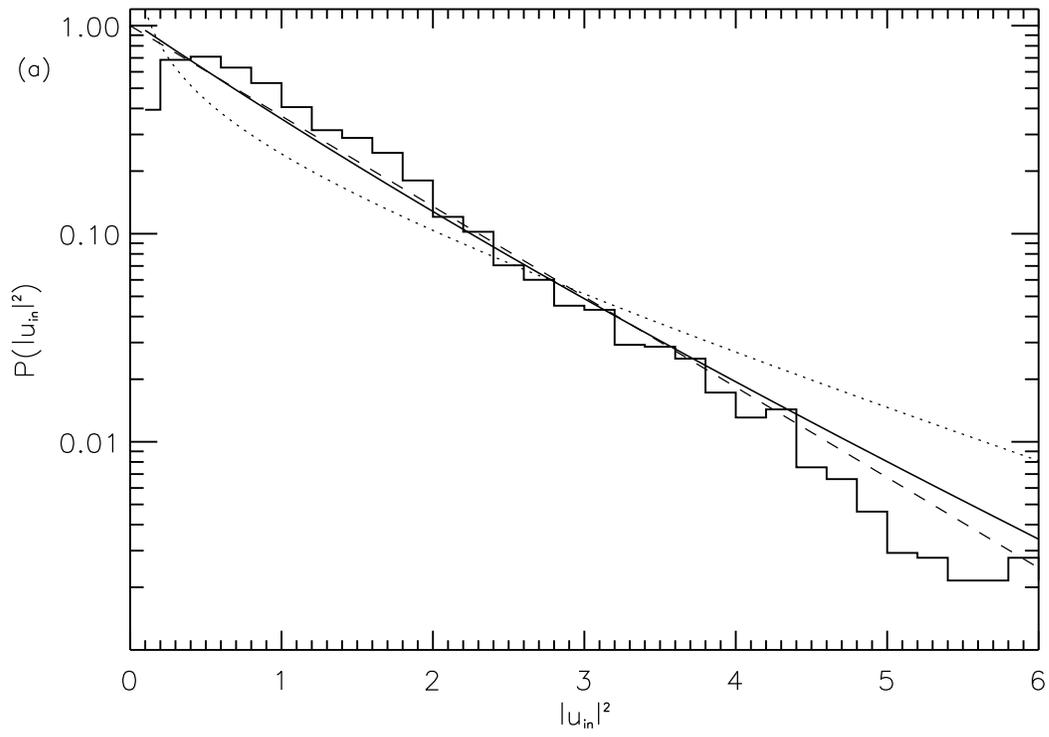
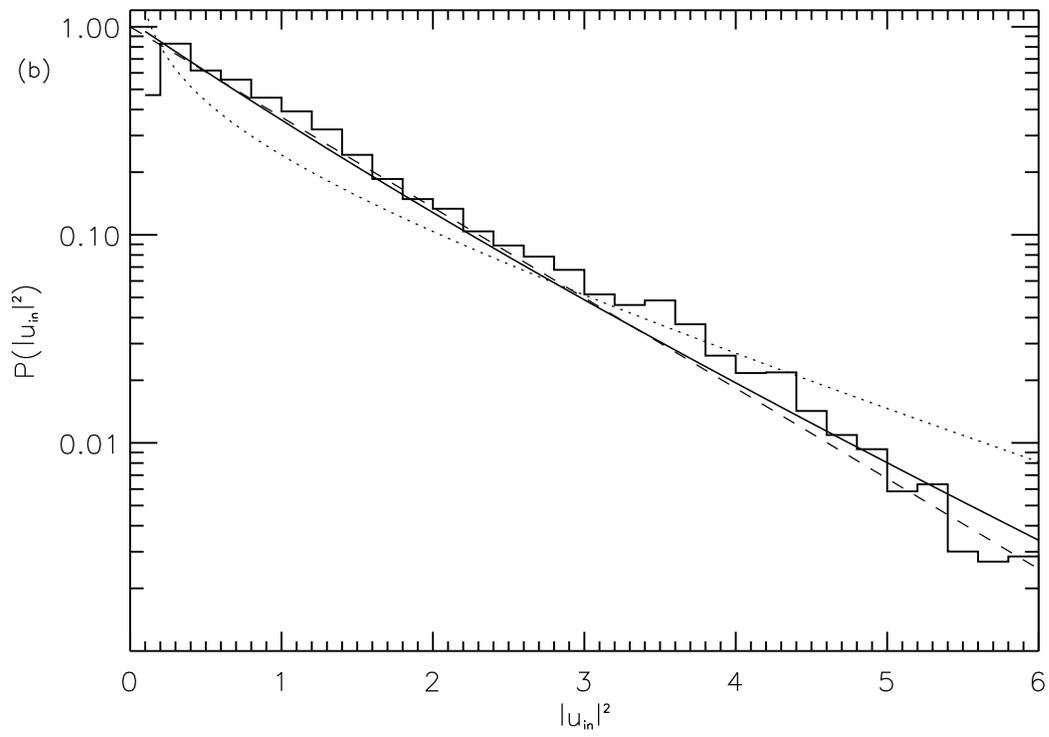
\begin{figure}
\caption{Histograms of $|u_{in}|^2$ for (a) reflection ($S_{11}$) and (b)
transmission ($S_{12}$) in a semilogarithmic plot. The dashed and dotted
lines correspond to Poisson distribution ($Z=1$, GUE) and
Porter-Thomas distribution ($Z\to\infty$, GOE), respectively. The solid line
gives the distribution (\ref{distribution2}) with the parameter $Z
\approx 1.06$ calculated from $<S_{11}>$.}
\label{high}
\end{figure}

In Fig.~\ref{high} the histograms obtained from the measured spectra for the
distribution of the internal wavefunction $|u_{in}|^2$ for (a) $S_{11}$
(reflection measurement) and (b) $S_{12}$ (transmission measurement) in the
region of overlapping resonances are shown. The results are compared with
the theoretical predictions (\ref{distribution2}) and (\ref{distribution})
with the parameter $Z\approx1.06$ calculated from the mean value $<S_{11}>$.
As we did not measure $S_{22}$ we assumed $<S_{11}>=<S_{22}>$ to calculate
$Z$ from Eq.~(\ref{z12}). Since the geometries of the antennae are
identical, this assumption seems to be plausible. In Fig.~\ref{gauss}c,d we
show the distributions of ${\rm Re}(S_{12})$ (c) and ${\rm Im}(S_{12})$ (d).
Again the Gaussian shapes are found in accordance with the theory.

The experiments have shown that the ansatz of random unitary matrices by
Pereyra and Mello \cite{per83} for the scattering matrix can perfectly
account for the observed distribution of $S$-matrix elements found in our
microwave billiard (apart from one point where imperfections in the
calibration make the comparison impossible). The change in the $Z$ parameter
from $Z=1.17$ in the case of separated resonances to $Z=1.06$ in the case of
overlapping  ones shows further that the presence of the 14 not used
antennae closed by 50 $\Omega$ transforms the system from the GOE to,
essentially, the GUE behaviour. In the region of overlapping resonances we
found transmission and reflection behaviour as being already essentially a
GUE-like (see Fig.~\ref{high}). In other words: the presence of the drains
in the form of closed antennae transforms the standing waves of the original
billiard, more or less completely, into running waves propagating from the
entrance antenna to the different exit ports.

{\acknowledgements} This research has been partially supported by the
Foundation for Theoretical Physics in Slemeno, Czech Republic and by the 
Deutsche Forschungsgemeinschaft via the SFB 185 Nichtlineare Dynamik. M.~K. 
acknowledges the support from Polish KBN Grant 2 P03B 093 09.

\end{document}